\documentclass[a4paper,11pt]{article}
\usepackage{jheppub} 

\usepackage[T1]{fontenc} 
 \usepackage{bm}
 \usepackage{ulem}
\newcommand{\Ds}{\displaystyle}                           
\newcommand{\be}{\begin{equation}}\newcommand{\ee}{\end{equation}}%
\newcommand{\ba}{\begin{eqnarray}}                        
 \newcommand{\ea}{\end{eqnarray}}                         
\def\MSbar{\relax\ifmmode\overline                        
           {\rm MS}\else{$\overline{\rm MS}${ }}\fi}     
  \def\ie{\hbox{\it i.e.}{ }} 
   \def\eg{\hbox{\it e.g.}{ }} 
\newcommand{\sss}[1]{\scriptscriptstyle{#1}}
\newcommand{\ssL}{{\mathcal L}}
\newcommand{\f}[2]{\frac{#1}{#2}}
\newcommand{\p}{\partial}
\newcommand{\ssst}[1]{\scriptscriptstyle{\text{#1}}}
\newcommand{\cf}{C_\text{F}}
\newcommand{\ca}{C_{\scriptscriptstyle{A}}}
\newcommand{\Ng}{N_{\scriptscriptstyle{A}}}
\newcommand{\cfi}{C_{\scriptscriptstyle{F,i}}}
\newcommand{\lb}{\left(}
\newcommand{\rb}{\right)}
\newcommand{\msbar}{$\overline{\text{MS}}$}
\newcommand{\gs}{g_{\scriptscriptstyle{s}}}
\newcommand{\dAAfNgex}{\frac{d_{\scriptscriptstyle{A}}^{abcd}d_{\scriptscriptstyle{A}}^{abcd}}{\Ng}}
\newcommand{\dFFfijNgex}{\frac{d_{\scriptscriptstyle{F,i}}^{abcd}d_{\scriptscriptstyle{F,j}}^{abcd}}{\Ng}}
\newcommand{\dFAfiNgex}{\frac{d_{\scriptscriptstyle{F,i}}^{abcd}d_{\scriptscriptstyle{A}}^{abcd}}{\Ng}}
\newcommand{\cfr}{C_{\scriptscriptstyle{F,r}}}
\newcommand{\tri}{T_{\scriptscriptstyle{F,i}}}
\newcommand{\trr}{T_{\scriptscriptstyle{F,r}}}
\newcommand{\dFr}{d_{\scriptscriptstyle{F,r}}}
\newcommand{\Nfi}{n_{\scriptscriptstyle{f,i}}}
\newcommand{\Nfj}{n_{\scriptscriptstyle{f,j}}}
\newcommand{\Nfr}{n_{\scriptscriptstyle{f,r}}}
\usepackage{color}
\definecolor{green}{rgb}{0.133,0.56,0}
\definecolor{DarkGreen}{rgb}{0.04,0.5,0.1}

\def\1{\hbox{{1}\kern-.25em\hbox{l}}}

\title{The $\{\beta\}$-expansion for Adler function, Bjorken Sum Rule, and the Crewther-Broadhurst-Kataev
 relation \\ at order $O(\alpha_s^4)$
 }
 \author[a]{P. A. Baikov,}
  \author[b]{ S.~V.~Mikhailov}
  \affiliation[a]{
Skobeltsyn Institute of Nuclear Physics, Lomonosov Moscow State University,
1(2), Leninskie gory, Moscow  119991, Russian Federation
        }
   \affiliation[b]{Bogoliubov Laboratory of Theoretical Physics, JINR,
                141980 Dubna, Russia}
\emailAdd{baikov@theory.sinp.msu.ru}
\emailAdd{mikhs@theor.jinr.ru}
\abstract{%
 We derive explicit expressions for the elements of the $\{ \beta \}$-expansion for
the nonsinglet Adler $D_A$-function and Bjorken polarized sum rules $S^{Bjp}$
in the  N$^4$LO using
recent results by Chetyrkin for these quantities computed within extended QCD including
any number of fermion representations.
We discuss the properties of the $\{ \beta \}$-expansion for $D_A$ and $S^{Bjp}$ at higher orders
which follow from the Crewther \cite{Crewther:1972kn} and the Broadhurst-Kataev \cite{Broadhurst:1993ru} relation.
}
\begin{document}
\maketitle
\section{Introduction}
\label{intro}
The knowledge of the detailed structure of the perturbation  QCD series (pQCD) is rather important
for a variety of problems of which the  optimization via the renormalization group (RG)
of the truncated series is the most known.
The detailed structure means  revealing the structure of the pQCD expansion
coefficients by means of the $\{\beta\}$-expansion \cite{Mikhailov:2004iq}
that looks now as a double series
rather than a usual one \cite{Mikhailov:2004iq,Kataev:2010du,Mikhailov:2016feh,Kataev:2016aib}.
We shall explore this structure  for the QCD renormalization group  invariant
(RGI) quantities (having no anomalous dimension)  depending on one-scale.
To obtain elements of perturbative coefficients within the $\{\beta\}$-expansion,
we apply the algebraic approach elaborated in \cite{Mikhailov:2016feh}.
Their knowledge allows us to pose the problem of optimizing the series \cite{Mikhailov:2004iq, Kataev:2014jba} and also
helps to relate the elements of different RGI quantities,
see \cite{Kataev:2010du,Kataev:2016aib,Kataev:2014jba,Mikhailov:2016feh}.
As  examples of these quantities we consider here the physically important  Adler function $D(Q^2,\mu^2)$
\ba
\label{DA}
\!\!\!\!D^{\rm EM}\left(\frac{Q^2}{\mu^2},a(\mu^2)\right)\!\! =\!\! \left(  \sum_i q_i^2
\right)\!\!
d_F D_{\rm NS}\left(\frac{Q^2}{\mu^2},a(\mu^2)\right)
\!+\! \left(\sum_i q_i \right)^2\!\! d_F D_{\rm S}\left(\frac{Q^2}{\mu^2},a(\mu^2)\right),
\ea
and Bjorken polarized sum rules $S^\text{Bjp}(Q^2,\mu^2)$,

\ba
&&S^\text{Bjp}(Q^2)
=\frac{1}6\,\bigg{|}\frac{g_A}{g_V}\bigg{|}~\left[
C^\text{Bjp}_{\rm NS}\left(\frac{Q^2}{\mu^2},a(\mu^2)\right) +
\left(\sum_{i} q_i \right )C^\text{Bjp}_{\rm S}\left(\frac{Q^2}{\mu^2},a(\mu^2)\right)\right]\,.
\ea
Here $q_i$ is the electric charge of the quark, $g_A, g_V$ -- nucleon
axial and vector charges, $d_F$ -- the dimension of the standard quark color representation,
strong coupling -- $a =\alpha_s/(4\pi)$.
The perturbation expansions for these quantities  are one of the most advanced, see Appendix \ref{App:B}.
Perturbative expressions for the nonsinglet (NS)
coefficient functions of both these quantities at the default choice of  scale $\mu^2=Q^2$ read 
\begin{eqnarray} \label{eq:PT-D-C}
  D_{\rm NS}(a(\mu^2)) = 1 + \sum_{n\geq1} a^n_s(\mu^2)~d_n, ~~~C^\text{Bjp}_{\rm NS}(a(\mu^2)) =    1 + \sum_{n\geq
  1}a^n(\mu^2) \ c_n.
 \end{eqnarray}
They  were obtained in order of $O(a^4)$ in the \msbar-scheme in \cite{Baikov:2010je,Baikov:2012zn}.
We will use here only the NS parts of these quantities,
 $D=D_\text{NS},~C^\text{Bjp}=C^\text{Bjp}_\text{NS}$, omitting
the notation NS further in the text.
The  expansion  coefficients $d_n$ and $c_n$ in Eqs.(\ref{eq:PT-D-C})  are the combinations
of the color coefficients only.
Let us recall the structure of these perturbation coefficients.
The $\{\beta\}$-expansion representation \cite{Mikhailov:2004iq}
 prescribes to decompose  $d_n$ or $c_n$, or coefficients of  any other of RGI quantity as 
 \begin{subequations}
\label{eq:d_beta}
\begin{eqnarray}
\label{eq:d_1}
a^0~~~1&&  \nonumber\\
a^1~~d_1&=&d_1[0]\, , \\
a^2~~d_2&=&\! \beta_0\,d_2[1]
  + d_2[0]\, ,\label{eq:d_2}\\
a^3~~  d_3&=&\!
  \beta_0^2\,d_3[2]
  + \beta_1\,d_3[0,1]
  +       \beta_0 \,  d_3[1]
  + d_3[0]\, ,\label{eq:d_3} \\
a^4~~  d_4
   &=&\! \beta_0^3\, d_4[3]
     + \beta_1\,\beta_0\,d_4[1,1]
     + \beta_2\, d_4[0,0,1]
     + \beta_0^2\,d_4[2]
     + \beta_1  d_4[0,1]
     + \beta_0\,d_4[1] \nonumber \\
   && \phantom{\beta_0^3\, d_4[3]+ \beta_1\,\beta_0\,d_4[1,1]+ \beta_0^2\,}~+d_4[0]\,,
       \label{eq:d_4} \\
 \ldots&& \ldots \,,    \nonumber
\end{eqnarray}
\end{subequations}
where  $\beta_i$ are the coefficients of the QCD $\beta$-function
\begin{equation}
\label{eq:beta}
\mu^2\frac{d a(\mu^2)}{d \mu^2}=
-\beta(a)=-a^{2}(\mu^2) \sum_{i\geq 0} \beta_{i}~ a^{i}(\mu^2)\,.
\end{equation}
Here we shall consider the results up to $O(a^4)$ in Eqs.(\ref{eq:d_beta}), while for the higher orders we
shall provide some predictions, \eg, for the elements of the next fifth order, see
the tail of the previous expansion in Eq.(\ref{eq:d_5}) and higher in Eq.(\ref{eq:d_n}) of Sec.\ref{sec:4}.
The designation $i_0, i_1,\ldots$ of the arguments of $d_n[i_0,i_1,\ldots]$ denotes
the exponents of the accompanying factor $\beta_0^{i_0} \beta_1^{i_1}\ldots$
The decompositions in Eqs.(\ref{eq:d_beta}) contain  complete knowledge
of the strong charge renormalization by  using there all  possible
$\beta_i$-terms in each order of expansion, for details see \cite{Mikhailov:2004iq,Kataev:2010du,Kataev:2014jba}.
This kind of expansion is the essential part of the procedures for the optimization of
the perturbation series, \eg, the decomposition (\ref{eq:d_2})
 was the starting point of the well-known BLM prescription
\cite{Brodsky:1982gc} in NLO.
Higher-order calculations are labor intensive and should be used to the maximum effect, \ie, be optimized.

At NLO of pQCD the decomposition in (\ref{eq:d_2}) looks quite evident, the term proportional
to $\Ds \frac{4}3 T_\text{R} n_f$ is an unambiguous attribute of
$\Ds \beta_0 = \frac{11}{3} C_A - \frac{4}{3} T_R n_f$
 in $d_2$ or other physical quantity.
 Whereas in NNLO QCD we can no longer separate the corresponding terms at
 $\beta_0$ and  $\beta_1$ in (\ref{eq:d_3}).
 To sharpen the problem, let us note that in pure gluodynamics, \ie, without quark loops, at $T_\text{R} n_f\equiv0$,
 it is impossible to carry out this decomposition even for the lowest case of $d_2$. 
 How to obtain  elements of the decomposition in higher orders?
We solve the problem introducing additional degrees of freedom (d.o.f.),
 new fields that interact following the universal gauge principle and
  enter only in intrinsic loops.
Using the fermions in the adjoint representation
(\eg, MSSM light gluino $\Ds \frac{C_{A}}2 n_{\tilde{g}}$)
as an additional d.o.f., a simple algebraic scheme to obtain the elements of the $\{\beta\}$-expansion
was formulated in \cite{Mikhailov:2016feh}.
Here we significantly extend this approach based on the results in \cite{Zoller:2016sgq,Chetyrkin:2017mwp,K:2022ebj}.
It should be emphasised that our approach doesn't use any specifics of $D$ or $C^{Bjp}$
and is universally applied to the expansion of any RGI quantities.

Another subject of our consideration is the Crewther relation \cite{Crewther:1972kn,Crewther:1997ux}
that determines the perturbation structure of the product $D\cdot C^\text{Bjp}$
that is inspired by the conformal symmetry breaking arguments, see Sec.\ref{sec:3}.
 This relation was significantly evolved later by D. Broadhurst and A. Kataev in \cite{Broadhurst:1993ru};
 therefore, we will name it the Crewther-Broadhurst-Kataev (CBK) relation.
 The general proof of the CBK relation has been discussed  in \cite{Braun:2003rp}.

 The usage of the $\{\beta\}$-expansion for CBK relation leads to predictions of
 mutual relations for the elements
 of $D$ and $C^\text{Bjp}$ for  any order of pQCD \cite{Kataev:2010du,Mikhailov:2016feh}.
These confirmations/predictions are further continued and elaborated in Sec.\ref{sec:3}-\ref{sec:4},
where we mention also other attempts to obtain the $\{\beta\}$-expansion.
Our main results are listed in Conclusion.
In Appendices A and B we have collected the important results for $\beta$-function, $D$, and $C^\text{Bjp}$
 obtained in \cite{Zoller:2016sgq,K:2022ebj}.
\section{The N$^4$LO $\{\beta\}$-expansion for the\\ Adler D-function and Bjorken polarized SR $C^\text{Bjp}$}
 \label{sec:2}
\subsection{How to decompose RG invariants}
 \label{sec:2.1}
The algebraic scheme, presented in \cite{Mikhailov:2016feh},
is well algorithmized  and appropriate to apply to high loop results.
The key role in the scheme plays the set of zeros of $\beta$-function coefficients
$ \beta_0(\{ R\}), \beta_1(\{ R\}), \beta_2(\{ R\}),\ldots $  and zeros of different sets of these $\beta_k$.
Here $\{ R\}$ means a set of fermion  degrees of freedom -- d.o.f.,
appearing in a QCD-like model extended to include
any number of different fermion representations of the gauge group, (QCDe), see \cite{K:2022ebj}
 and Appendix \ref{App:A1}.
The more d.o.f. we include in consideration,
the more higher order expansion coefficients  $ d_n $
can be untangled with respect to $ d_n [.]$.
Let us illustrate this with an example of the $d_4$ decomposition.
Suppose we have the expressions for $d_4(x_0,x_1,x_2)$ and for the coefficients $\beta_{0,1,2}(x_0,x_1,x_2)$
that depend on the attributes of d.o.f. $\{x_0,x_1,x_2\}$.
Suppose we find a root $(x_{0,0},x_{1,0},x_{2,0})\!:\!\{\beta_{0,1,2}(x_{0,0},x_{1,0},x_{2,0})\!=\!0\}$.
So, following  Eq.(\ref{eq:d_4}), we obtain $d_4(x_{0,0},x_{1,0},x_{2,0})\!=\!d_4[0]$.
Taking the  reduced condition $\{ \beta_{0,1}(x_{0,0},x_{1,0})\!=\!0\}$ for the roots,
one keeps two terms in $d_4$,
$d_4(x_{0,0},x_{1,0},x_{2})\!=\!d_4[0]+ \beta_2(x_{0,0},x_{1,0},x_{2}) d_4[0,0,1]$,
from which one can extract $d_4[0,0,1]$, and so on.
Directly following the scheme presented in \cite{Mikhailov:2016feh}, Sec.3.3 there, and
using the explicit expressions for $D\left(\{ R\}\right)$ and $C^\text{Bjp}\left(\{ R\}\right)$ that involve
new fermions d.o.f. presented in Appendices \ref{App:A}, \ref{App:B},
one can obtain expressions for all of the elements $d_4[.]$ and $c_4[.]$.
Let us emphasize, we need new d.o.f. \textit{only to perform the decomposition},
 after that we return from  QCDe to the standard  QCD, $\{R \} \rightarrow T_\text{R}n_f$.
The trace of the general gauge principle, presented  first via interactions of different d.o.f.
within QCDe, ends up as a structure of the  $\{\beta\}$-expansion. \\
The results for the $\{\beta\}$-expansion presented below for $D$ in $O(a^3)$
were first obtained in \cite{Mikhailov:2004iq} using calculations with light MSSM gluino (one new d.o.f.)
\cite{Chetyrkin:1996ez,Clavelli:1996pz}.
In this case, the set $\{R \}$ is reduced to the appearance of a pair of attributes $T_{R}n_f,~\frac{C_{A}}2
n_{\tilde{g}}$ in the final expressions.
The remaining elements for $C^\text{Bjp}$ in this order were first restored in \cite{Kataev:2014jba} using the CBK
relation.
\subsection{Decomposition for the Adler D-function}
\label{sec:2.2}
For the Adler function $D$ the corresponding elements in order $O(a^3)$ read
\cite{Mikhailov:2004iq,Kataev:2010du,Kataev:2014jba}
\begin{subequations}
\label{eq:d1-4}
 \begin{eqnarray}
d_1&=&3{\rm C_F}; \label{D-11}\\
d_2[1]&=&d_1\left(\frac{11}2-4\zeta_3\right);~
d_2[0]=d_1\left(\frac{\rm C_A}3-\frac{\rm C_F}2\right); \label{D-21} \\
d_3[2]&=&d_1\left(\frac{302}9-\frac{76}3\zeta_3\right);~d_3[0,1]=d_1\left(\frac{101}{12}-8\zeta_3\right);\label{D-32}\\
d_3[1]&=&
    d_1\left[{\rm C_A}\left(-\frac{3}4 + \frac{80}3\zeta_3 -\frac{40}3\zeta_5\right) -
    {\rm C_F}\left(18 + 52\zeta_3 - 80\zeta_5\right) \label{D-31}\right]; \label{eq:d31} \\
d_3[0]&=& d_1\Bigg[{\rm C_A^2}\left(\frac{523}{36}- 72 \zeta_3\right)
    +\frac{71}3 {\rm C_A C_F} - \frac{23}{2} {\rm C_F^2}\Bigg]~.  \label{D-30}
\end{eqnarray}
\end{subequations}
The results in Eqs.(\ref{eq:d1-4}) were confirmed based on the results obtained within QCDe \cite{K:2022ebj}
and briefly outlined in Appendices \ref{App:A} and \ref{App:B}.
Besides these results has been analyzed and discussed in \cite{Ma:2015dxa} from the point of view of PMC approach.
Then we have obtained the $\{ \beta\}$-expansion elements in order of $O(a^4)$,
\begin{subequations}
 \label{eq:d_4expr}
 \begin{eqnarray}
 d_4[3] &=& C_\text{F}\left(\frac{6131}9 - 406 \zeta_3 - 180 \zeta_5\right); \label{eq:d_43}\\
  d_4[1,1]&=& C_\text{F}\left(385 - \frac{1940}3 \zeta_3+ 144 \zeta_3^2 + 220 \zeta_5\right);\label{eq:d_411} \\
 d_4[2] &=&-C_\text{F} \left[ C_\text{F} \left(\frac{6733}{8} + 1920 \zeta_3 - 3000 \zeta_5\right) + \right.\nonumber
 \\
         &&\left. \phantom{-C_\text{F}\Big[ }C_\text{A} \left(\frac{20929}{144} - \frac{12151}{6} \zeta_3 + 792
         \zeta_3^2
         + 1050 \zeta_5\right)\right]; \label{eq:d_42}\\
  d_4[0,0,1]&=& C_\text{F} \left(\frac{355}{6} + 136 \zeta_3 - 240 \zeta_5\right); \label{eq:d_4001}
 \end{eqnarray}
 \begin{eqnarray}
 d_4[1] &=& C_\text{F}\bigg[
 -C_\text{F}^2 \left(\frac{447}2 - 42 \zeta_3 - 4920 \zeta_5 + 5040 \zeta_7\right)+\nonumber\\
 &&\phantom{C_\text{F} \bigg[ } C_\text{A} C_\text{F} \left(\frac{3301}4 - 678 \zeta_3 - 2280 \zeta_5 + 2520
 \zeta_7\right)+\nonumber\\
 &&\phantom{C_\text{F} \bigg[ }C_\text{A}^2 \left(\frac{16373}{36} - \frac{17513}{3} \zeta_3 + 2592 \zeta_3^2 +
    3030 \zeta_5 - 420 \zeta_7\right)\bigg] \label{eq:d_41}; \\
  d_4[0,1]&=& - C_\text{F} \left[C_\text{A} \left(\frac{139}{12} + \frac{1054}3 \zeta_3 - 460 \zeta_5\right) +
    C_\text{F} \left(\frac{251}4 + 144 \zeta_3 - 240 \zeta_5 \right)\right]; \label{eq:d_401}
 \end{eqnarray}
 \begin{eqnarray}
 d_4[0] &=&\tilde{d}_4[0]+ \delta d_4 \nonumber \\
         &=&C_\text{F}^4 \left(\frac{4157}{8} + 96 \zeta_3\right) -
 C_\text{A} C_\text{F}^3 \left(\frac{2409}{2} + 432 \zeta_3\right)+
 C_\text{A}^2 C_\text{F}^2 \left(\frac{3105}{4} + 648 \zeta_3\right) + \nonumber \\
 &&C_\text{A}^3 C_\text{F} \left(\frac{68047}{48} + \frac{8113}{2} \zeta_3 - 7110 \zeta_5\right) + \delta d_4\,;
 \label{eq:d_40}
 \end{eqnarray}
 \begin{eqnarray}
 \delta d_4&=&-16\Big[n_f \frac{d_F^{a b c d}d_F^{a b c d}}{d_F} \left(13 + 16 \zeta_3 - 40 \zeta_5\right) +
 \frac{d_A^{a b c d}d_F^{a b c d}}{d_F} \left(-3 + 4 \zeta_3 + 20\zeta_5 \right)\Big]\,. \label{eq:d_40delta}
\end{eqnarray}
 \end{subequations}
Equation (\ref{eq:d_43}) for $d_4[3]$ should be compared with the term at $C_\text{F}T_f^3$ in Eq.(10) in
\cite{Baikov:2010je}
and even with the early general formulae for renormalon chain contributions obtained in
\cite{Broadhurst:1993ru,Broadhurst:1992si}.
\subsection{Decomposition for the Bjorken SR $C^\text{Bjp}$}
 \label{sec:2.3}
Here we confirm the results for $c_{2-3}[.]$ up to the order $O(a^3)$ \cite{Kataev:2014jba}
\textit{through direct calculations}
not using the CBK relation (following CBK this was done in \cite{Kataev:2014jba}),
  \begin{subequations}
 \label{eq:c1-4}
\begin{eqnarray}
c_1&=&-3~{\rm  C_F}; \label{c-11}\\
c_2[1]&=& 2 c_1;
c_2[0]= \left(\frac{1}{3}{\rm C_A}-\frac{7}{2}{\rm C_F}\right) c_1; 
 \label{c-21} \\
c_3[2]&=& \frac{115}{18} c_1;~c_3[0,1]=c_1\bigg(\frac{59}{12}-4\zeta_3\bigg);
\label{C-32} \\
c_3[1]&=& -c_1\bigg[{\rm C_F}\bigg(\frac{166}{9}- \frac{16}3\zeta_3\bigg) +
{\rm C_A}\bigg(\frac{215}{36}- 32 \zeta_3+\frac{40}{3}\zeta_5\bigg)\bigg];
\label{C-31} \\
c_3[0] &=&c_1\bigg[{\rm C_A^2}
\bigg(\frac{523}{36} - 72\zeta_3\bigg)+\frac{65}{3}{\rm C_F C_A}+ \frac{\rm C_F^2}{2}\bigg]\,.
\label{C-30}
\end{eqnarray}
\end{subequations}
The  elements $c_4[.]$ in order $O(a^4)$ are obtained based on the results
for $C^\text{Bjp}(\{R\})$ \cite{K:2022ebj} presented in  Appendices \ref{App:A} and \ref{App:B} following
the same universal procedure as it was applied to the $D$ case in Sec.\ref{sec:2.2},

\begin{subequations}
 \label{eq:c_4expr}
 \begin{eqnarray}
 c_4[3] &=& -C_\text{F}\frac{605}9; \\
  c_4[1,1]&=& C_\text{F}\left(-\frac{715}8+ \frac{677}3 \zeta_3 -220 \zeta_5\right); \\
 c_4[2] &=&C_\text{F} \left[ C_\text{F} \left(\frac{8057}{24} - 96  \zeta_3\right) +
C_\text{A} \left(\frac{28615}{144} - \frac{4105}6  \zeta_3 - 24  \zeta_3^2 + 370  \zeta_5\right)\right];    \\
  c_4[0,0,1]&=& C_\text{F} \left(-\frac{146}3 - 148 \zeta_3 + 240 \zeta_5\right); \\
 c_4[1] &=&C_\text{F} \bigg[C_\text{F}^2\left(-\frac{1478}3-824 \zeta_3+1520\zeta_5\right)- \nonumber  \\
 &&\phantom{C_\text{F} \bigg[ }C_\text{A} C_\text{F} \left(\frac{1177}{12} - \frac{5888}3 \zeta_3 + \frac{2000}{3}
 \zeta_5 + 1680 \zeta_7\right)- \nonumber  \\
 &&\phantom{C_\text{F} \bigg[ } C_\text{A}^2 \left(\frac{3829}{72}-2286 \zeta_3 + \frac{6250}{3} \zeta_5 - 420
 \zeta_7\right)\bigg]; \\
  c_4[0,1]&=&  C_\text{F} \left[C_\text{A} \left(\frac{109}4 + \frac{1006}{3} \zeta_3 - 460 \zeta_5\right) +
    C_\text{F} \left(\frac{1399}{12} -100 \zeta_3 \right)\right];
 \end{eqnarray}
 \begin{eqnarray}
 c_4[0] &=&\tilde{c}_4[0]+ \delta c_4 \nonumber \\
        &=& - C_\text{F}^4 \left(\frac{4823}{8} + 96 \zeta_3\right)
 + C_\text{A} C_\text{F}^3 \left(\frac{3201}{2} + 432 \zeta_3\right)-
 C_\text{A}^2 C_\text{F}^2\left(\frac{2055}{4} + 1944 \zeta_3\right)- \nonumber\\
 && C_\text{A}^3 C_\text{F} \left(\frac{68047}{48} + \frac{8113}{2} \zeta_3 - 7110 \zeta_5\right) + \delta c_4\,;
 \label{eq:c_40} \\
 \delta c_4=-\delta d_4&\!=\!&\!16\Big[n_f \frac{d_F^{a b c d}d_F^{a b c d}}{d_F} (13 + 16 \zeta_3 - 40 \zeta_5) +
 \frac{d_A^{a b c d}d_F^{a b c d}}{d_F} (-3 + 4 \zeta_3 + 20\zeta_5)\!\Big]. \label{eq:c_40delta}
\end{eqnarray}
 \end{subequations}
Let us remark that we first face here the d.o.f. dependence of the elements  $d_n[.]$,
see the terms $n_f$ in $c_4[0]$, (\ref{eq:c_40delta}), and $d_4[0]$, (\ref{eq:d_40delta}).
This kind of $n_f$ dependence is related to the special contributions from the  ``box'' subgraph rather than the
$a$-renormalization.
These terms $ \delta c_4=-\delta d_4$  were  obtained within the total expressions for
$c_4[0], d_4[0]$ in \cite{Cvetic:2016rot}, they were independently extracted in the form of
Eq.(\ref{eq:c_40delta}) in \cite{Mikhailov:2016feh}, see Appendix A therein. 
The other part of this ``box'' subgraph contributes to the charge renormalization (see, \eg, \cite{grozin:2007}).
The $ \delta c_4, \delta d_4$ are the single elements at $O(a^4)$ order depended on the d.o.f.
from the extended QCDe along with the $\beta$-function coefficients $\beta_{0-2}$.
All other elements $c_i[.], d_i[.]$ at $i\leqslant 4$ are universal,
while the effects of the  inclusion of these d.o.f.
are accumulated only in the $\beta$-function coefficients appearing in front of them.

\section{The constraints on $C^\text{Bjp}$ and $D_A$ elements from the CBK relation}
 \label{sec:3}
\subsection{The structure of the CBK relation}
\label{sec:3.1}
The $\{\beta\}$-expansion for the quantities $D, C^\text{Bjp}$ can serve for effective verification
of the CBK relation (CBKR) \cite{Kataev:2010du,Broadhurst:1993ru},
\begin{subequations}
 \label{eq:CBKR}
 \begin{eqnarray}
\label{CRe1}
 D(a) \cdot C^\text{Bjp}(a) &=& \1 + \beta(a)\,\cdot K(a)\,, \\
                                &&K(a)=\sum_{n=1} a^{n-1}\, K_n, \label{Cre2}
\end{eqnarray}
 \end{subequations}
where $K_n=K_n(\beta_0, \beta_1,\ldots)$ in Eq.(\ref{Cre2}) are polynomials in $\beta_i$.
The structure of the RHS of Eq.(\ref{CRe1}) can be traced from the evident representation for the LHS,
 \begin{eqnarray} \label{eq:DC}
 D(a) \cdot C^\text{Bjp}(a) &=&1+ \sum_{n\geqslant2} a^n \Big[d_n +c_n + \sum_{i \geqslant1}^{n-1}d_ic_{n-i}\Big]\,,
\end{eqnarray}
when the condition for the coefficients $c_1=-d_1$ is explicitly satisfied, see Appendix \ref{App:B}.
Further, one should compare the second terms in the RHS of Eq.(\ref{CRe1}) and Eq.(\ref{eq:DC}).
It is clear that the coefficients $K_n$ should have the same
$\{\beta\}$-structure, Eq.(\ref{eq:d_beta}),  as for $d_k~(c_k)$ but after factorization of the common $\beta(a)$.
We show these expansions for $K_{1,2,3}$ that are sufficient up to $O(a^4)$,
\begin{subequations}
 \label{eq:K_i}
 \begin{eqnarray}
K_1 &=& K_1[1], \label{eq:K_1} \\
K_2 &=& K_2[1] + \beta_0 K_2[2], \\
K_3 &=& K_3[1]+ \beta_0 K_3[2] + \beta_0^2 K_3[3]+ \beta_1 K_3[1,1]\,.
\end{eqnarray}
Keep in mind that the meaning of the arguments of $K_n[i_0,i_1,\ldots]$ is a bit different from that  for the original
elements $d_n[.]$;
 it denotes the exponents \textit{of all the accompanying}
$\beta_0^{i_0},\beta_1^{i_1},\ldots$, including  also those that appear in the common factor $\beta(a)$
in the RHS of (\ref{CRe1}).
For future considerations let us present the decomposition also for the following $K_4$ term,
 \begin{eqnarray}
K_4 &=& K_4[1]+ \beta_0 K_4[2] + \beta_0^2 K_4[3]+  \beta_0^3 K_4[4]+  \beta_2 K_4[1,0,1]+  \beta_1\beta_0^2 K_4[3,1]+
\nonumber\\
&&\beta_1 K_4[0,2] + \beta_1 K_4[1,1]+\beta_1\beta_0 K_4[2,1]\,.
\end{eqnarray}
 \end{subequations}
The elements $K_n[.]$ of the $\{\beta\}$-expansion help to get rid of the traces of fermion d.o.f.
 that will be absorbed into the coefficients $\beta_i$.
The factorization of the common $\beta(a)$, being put in base, allows one
to relate the different elements of the $\{\beta\}$-expansion of $d_4 (d_n)$
to the corresponding  elements $c_4 (c_n)$ and vice versa.

\subsection{The ``conformal'' \1 of the CBK relation}
\label{sec:3.2}
In the conformal limit
the coefficients $\beta_i=0$ and  the CBKR (\ref{CRe1}) returns to its initial form
 \cite{Crewther:1972kn} with only $\1$ in the RHS that shows the restoration of conformal symmetry.
In this case, one gets the reduced factors $D \to D_0,  C^\text{Bjp} \to C^\text{Bjp}_0$
of the product in Eq.(\ref{CRe1}) (or in Eq.(\ref{eq:DC})); so the corresponding series
$ D_0(a)$ and $ C^\text{Bjp}_0(a)$ are  inverse.
The condition $C^\text{Bjp}_0 \cdot D_0=\1$ is related to the $d_n[0], c_n[0]$ elements in every order (see
Eq.(2.8) in \cite{Mikhailov:2016feh} and \cite{Kataev:2014jba}),
\ba
c_n[0]+d_n[0] &=& - \sum_{l=1}^{n-1} d_l[0] c_{n-l}[0]\,, \label{eq:CI-PT0}
\ea
and leads to an explicit closed solution  with respect to $c_k[0]$ and vice versa ($c\leftrightarrows d$)

\ba \label{eq:CI-PT0n}
c_k[0]= (-)^k \text{det}[D^{(k)}_0] \equiv (-)^k \left|
 \begin{array}{lllccc}
                 d_1 & 1    & 0   &            &\ldots & 0 \\
                 d_2 & d_1  & 1   &            &\ldots & 0 \\
                 d_3 & d_2  & d_1 &            &\ldots& 0  \\
              \ldots &      &     &            &\ldots& 0  \\
              d_{k-1}&\ldots&\ldots&            & d_1  & 1  \\
                 d_k & d_{k-1} & d_{k-2}&\ldots&d_2 & d_1   \\
               \end{array}
             \right|\,.
\ea
Here $D^{(k)}_0$ is the matrix that consists of only $d_i\equiv d_i[0]$ elements.
 The general relation (\ref{eq:CI-PT0n}) can be  treated as a prediction
for $C^\text{Bjp}$ by means of $D $ (and vice versa) that is based on the  $\{\beta\}$-expansion and the conformal part
of CBKR.
In the fourth  order of $a$,  relation (\ref{eq:CI-PT0n}) reduces to
\begin{subequations}
 \begin{eqnarray}
  \label{eq:c4+d4}
 \underline{d_4[0]+c_4[0]}&=& \tilde{d}_4[0]+\tilde{c}_4[0]= 2d_1 d_3[0]-3d_1^2d_2[0]+d_2[0]^2+d_1^4 \\
              &=&3{\rm C^2_F}\left[ 132 {\rm C_F }{\rm C_A} -\frac{111}{4}{\rm C^2_F}+\left(\frac{175}2-432
              \zeta_3\right)
              {\rm C^2_A}\right]\,,
 \end{eqnarray}
  \end{subequations}
 that is fulfilled \textit{automatically} (underlined terms), as well as all the previous ``zero'' element's sums
 $d_i[0]+c_i[0]$ at $i \leqslant 4$.
In the LHS  of Eq.(\ref{eq:c4+d4}) we use the components from Eqs.(\ref{eq:d_40}, \ref{eq:c_40}), while in the RHS --
the
components $d_i[0]$
from Eq.(\ref{eq:d1-4}).
The result for the sum in the LHS of Eq.(\ref{eq:c4+d4}) was already predicted in \cite{Kataev:2010du} (Eq.(21) there)
and in  \cite{Mikhailov:2016feh}, see Appendix A therein.
In the following fifth order the relation (\ref{eq:CI-PT0n}) leads to the prediction for
 the sum in the LHS (doubly underlined terms) obtained in the fourth order in the RHS,
\begin{subequations}
 \begin{eqnarray}
 \label{eq:c5+d5}
 \underline{\underline{d_5[0]+c_5[0]}} &=& 2 d_1 d_4[0]+ 2 d_3[0] d_2[0] - 3 d_3[0] d_1^2 + 4 d_2[0]d_1^3 - 3 d_2^2[0]
 d_1 -d_1^5 \\
               &=& d_1 \Bigg[ C_\text{A}^2 C_\text{F}^2\, 27 \left(43 + 128 \zeta_3\right) +  C_\text{F}^4
               \left(\frac{2485}2 + 192 \zeta_3\right) - C_\text{A} C_\text{F}^3 \left(3097 + 864 \zeta_3\right) +
               \nonumber \\
   &&\phantom{d_1 \Bigg[ } C_\text{A}^3 C_\text{F} \left(\frac{206233}{72} + 7969 \zeta_3 - 14220 \zeta_5\right)+2
   \delta
   d_4 \Bigg]\,. \label{eq:c5+d5-b}
 \end{eqnarray}
  \end{subequations}
  In general, see \cite{Mikhailov:2016feh},  if the elements $d_k[0]$ are known up to an order $(n-1)$,
  then the sum $d_n[0]+c_n[0]$ in order $n$ is

\ba
c_n[0] + d_n[0] =  (-)^n \left|
\begin{array}{lllccc}
                 d_1 & 1    & 0   &            &\ldots & 0 \\
                 d_2 & d_1  & 1   &            &\ldots & 0 \\
                 d_3 & d_2  & d_1 &            &\ldots& 0  \\
              \ldots &      &     &            &\ldots& 0  \\
              d_{n-1}&\ldots&\ldots&            & d_1  & 1  \\
                 \bm{0} & d_{n-1} & d_{n-2}&\ldots&d_2 & d_1   \\
               \end{array}
             \right|\,, \label{eq:CI-PTsum}
\ea
or at $c\leftrightarrows d$, as it follows from (\ref{eq:CI-PT0n}).

\subsection{The conformal symmetry-breaking term of the CBK relation}
\label{sec:3.3}
On the other hand, one can apply in the RHS of Eq.(\ref{CRe1}) the second term proportional to $\beta(a)$
that expresses the break of conformal symmetry.
The factorization of the whole $\beta(a)$  leads to the chains of equalities
\cite{Kataev:2010du,Kataev:2014jba} for the first terms $K_n[1]$ in the representation (\ref{eq:K_i}).
These equalities can be easily traced from the terms under the sum in the RHS of Eq.(\ref{eq:DC}).
Really, for such factorization the factors appearing in front of the successive coefficients $\beta_i$
should be equal to one another for any  position of $i$ -- the  argument of the factor.
For the combination $d_n+c_n$ in the sum in (\ref{eq:DC}) it leads
\begin{subequations}
 \begin{eqnarray}
  \label{eq:c2_1+d2_1}
\!\!\!\!\!\!\!\!\! \beta_0&&~~~~~~~\beta_1 \hspace*{15mm}\beta_2 \hspace*{6mm} \ldots \nonumber\\
\!\!\!\!\!\!\!\!\! K_1[1] &=&K_2[0,1]= K_3[0,0,1]\!=\ldots=\nonumber \\
\!c_2[1]+d_2[1]\!&=&\!c_3[0,1]+d_3[0,1]\!=\!\! \underline{c_4[0,0,1]+d_4[0,0,1]} \!\!=\!\!
3C_\text{F}\! \left(\frac{7}2-4\zeta_3 \right) \label{eq:c4__1+d4__1} \\
\!\! &=&\!\! \underline{\underline{c_5[0,0,0,1]+d_5[0,0,0,1]}} \!=\! c_n[\underbrace{0,0,\ldots, 1}_{n-1}] +
d_n[\underbrace{0,0,\ldots, 1}_{n-1}].\label{eq:cn_1+dn_1}
 \end{eqnarray}
   \end{subequations}
Then, the  factorization of the linear $\beta_i$  terms   leads to equalities for the subleading
corrections $K_{2,3,4,\ldots}[1]$ ( see the first terms in the RHS of Eq.(\ref{eq:K_i})),

\begin{subequations}
 \begin{eqnarray}
 K_2[1]&=&K_3[0,1]= K_4[0,0,1]\!=\ldots=\nonumber  \\
       &=&c_3[1] + d_3[1] + d_1 (c_2 [1] - d_2 [1]) =  \nonumber \\
&=& \underline{ c_4[0,1] + d_4[0,1]} + d_1(c_3[0,1] - d_3[0,1]) =   \nonumber \\
&=& {\rm  C_F^2}\bigg(-\frac{397}{6}-136\zeta_3+240\zeta_5\bigg)+{\rm  C_F C_A}\bigg(\frac{47}3-16\zeta_3\bigg)
\label{eq:c3_1+d3_1}\\
&=& \underline{\underline{c_5[0,0,1] + d_5[0,0,1]}} + d_1( c_4[0,0,1] - d_4[0,0,1])=\dots \nonumber\\
&=&c_n[\underbrace{0,\ldots,1}_{n-2}] +d_n[\underbrace{0,\ldots,1}_{n-2}]
 + d_1 ( c_{n-1} [\underbrace{0,\ldots, 1}_{n-2} ] -
d_{n-1}[\underbrace{0,\ldots, 1}_{n-2} ]).
\label{eq:cn_01+dn_01}
\end{eqnarray}
 \end{subequations}
Equations (\ref{eq:c2_1+d2_1},\ref{eq:c3_1+d3_1}) are fulfiled automatically,
 the current $O(a^4)$ results are checked and underlined;
 see the corresponding elements in Eq.(\ref{eq:d_4expr}) and Eq.(\ref{eq:c_4expr}).
Equations (\ref{eq:cn_1+dn_1},\ref{eq:cn_01+dn_01}) can serve as natural predictions
   for any highest orders, the next order $O(a^5)$ results are doubly underlined.
Following this line, Eq.(\ref{eq:c4_1+d4_1}) provides  predictions for the $ c_5[0,1]+d_5[0,1]$ for 6-loop results
in the RHS of Eq.(\ref{eq:P^3_1}) and further,
\begin{subequations}
 \label{eq:3.9}
 \begin{eqnarray}
 K_3[1]&=&K_4[0,1]=\ldots=\nonumber  \\
       &=&c_4[1]+d_4[1]+d_1\big(c_3[1]-d_3[1]\big)+d_2[0]c_2[1]+d_2[1]c_2[0]   \nonumber \\
 &=&C_\text{A}^2 C_\text{F}\! \left(\frac{3213}8 - \frac{10655}3 \zeta_3 + 2592 \zeta_3^2 + \frac{2840}3 \zeta_5\right)
- \nonumber \\
&& C_\text{F}^3\! \left(\frac{2471}{12} - 488 \zeta_3 + 5720 \zeta_5 - 5040 \zeta_7\right) + \nonumber \\
&& C_\text{A} C_\text{F}^2 \left(\frac{4591}6 + \frac{2306}3 \zeta_3 - \frac{8120}3 \zeta_5 + 840 \zeta_7\right)
\label{eq:c4_1+d4_1}
 \end{eqnarray}
  \begin{eqnarray}
&=&\!\! c\underline{\underline{_5[0,1]+d_5[0,1]}}+d_1
\big(c_4[0,1]-d_4[0,1]\big)\!+\!d_2[0]c_3[0,1]\!+\!c_2[0]d_3[0,1]=\!\ldots \label{eq:P^3_1} \\
&=&\!\! c_{n+1}[\underbrace{0,\ldots,1}_{n-2}] +d_{n+1}[\underbrace{0,\ldots,1}_{n-2}]
 + d_1 \Big( c_{n} [\underbrace{0,\ldots, 1}_{n-2} ] -
d_{n}[\underbrace{0,\ldots, 1}_{n-2} ] \Big)+ \nonumber \\
&& d_2[0]\,c_{n-1}[\underbrace{0,\ldots, 1}_{n-2} ]+ c_2[0]\,d_{n-1}[\underbrace{0,\ldots, 1}_{n-2} ]\,.
\label{eq:cn_001+dn_001}
 \end{eqnarray}
  \end{subequations}
Similarly, the next subleading $K_4[1]$ and its chain of equations can be constructed.
We present its expression here for future results
and also because it contains the last term linear in $\beta_i$
from Eq.(\ref{eq:d_4expr},\ref{eq:c_4expr}),
\begin{eqnarray}
 K_4[1]&=& c_5[1]+d_5[1]+d_1
\big( c_4[1]-d_4[1] \big)+\left(d_2[0]c_3[1]+d_3[1]c_2[0]\right)([1]\leftrightarrow[0])\,, \label{eq:3.10}
 \end{eqnarray}
 while in the general case,  we obtain
 \begin{eqnarray}
\!\!\!\!\!\!\!\!\! K_n[1]\!&\!=\!&\! c_{n+1}[1]\!+\!d_{n+1}[1]+d_1
\big(c_n[1]-d_n[1] \big)\!+\! \sum_{k=2}^{n-1}\left(d_k[0]c_{n+1-k}[1]+c_k[0]d_{n+1-k}[1] \right). \label{eq:3.12}
 \end{eqnarray}
These examples demonstrate that the elements of the $\{\beta\}$-expansion provide
\textit{appropriate building blocks} to analyse and construct the CBK relation.
 Following this way we  exhausted all the cases linear in $\beta_i$.

Let us consider now the content of the contributions  to $K_{2,3}$ that are higher in exponents of $\beta_i$.
Taking in the RHS of Eq.(\ref{eq:DC}) the $\{\beta \}$-expansion for the  terms $c_n$ and $d_i$,
we come to obvious expressions for the elements $K_{2,3}[.]$,
 \begin{subequations}
  \label{eq:K2-3}
 \begin{eqnarray}
&K_2[2]&= \dashuline{c_3[2]+d_3[2]}  = C_\text{F}\! \left(\frac{163}{2}-76\zeta_3 \right)\,, \label{eq:cn_2+dn_2}\\
&K_3[1,1]&= c_4[1,1]+d_4[1,1]=C_\text{F}\! \left(\frac{2365}{8}-421\zeta_3+144\zeta_3^2 \right),
\end{eqnarray}
 \begin{eqnarray}
&K_3[2]&= c_4[2]+d_4[2]+\dashuline{d_1\left(c_3[2]-d_3[2] \right)+d_2[1] c_2[1]} \nonumber \\
      &&= C_\text{F}\! \bigg[C_\text{A} \left(\frac{427}{8}+1341\zeta_3-816\zeta_3^2-680\zeta_5 \right)\nonumber \\
      &&\phantom{= C_\text{F}\! \bigg[} -C_\text{F}\left(\frac{11573}{12} + 1716 \zeta_3 -3000 \zeta_5\right)\bigg]\,,
      \\
&K_3[3]&= \dashuline{c_4[3]+d_4[3]}=C_\text{F}\! \left(614-406\zeta_3-180\zeta_5 \right).
\end{eqnarray}
 \end{subequations}
For every  second equality in Eqs.(\ref{eq:K2-3}) we have used explicit results from
Eqs.(\ref{eq:d1-4}, \ref{eq:d_4expr})
for $d_i[.]$ and from Eqs.(\ref{eq:c1-4}, \ref{eq:c_4expr}) for $c_i[.]$.

 It is easy to represent those parts of $K_n[i]$ that are generating by the renormalon chain contributions  possessed
 the
 maximum powers of $ \left(a \beta_0\right)^n$ in each order.
 The elements $d_n[n-1]$ and $c_n[n-1]$ are known explicitly from the results in \cite{Broadhurst:1993ru,Broadhurst:1992si}
 and one does not need new d.o.f. for it.
 The corresponding parts, \eg, marked with dashed lines in (\ref{eq:K2-3}), can be obtained from the expression for
 $K(a)$, (see Eq.(\ref{sec:3.1}), at $n \geqslant 2$), taken in order of $O(a^n)$
 \begin{eqnarray} \label{eq:renormalon}
\hspace*{-10mm}\beta(a)\cdot K(a)&\Rightarrow&\!\! a^2 \beta_0 \cdot \bigg\{ \left(a \beta_0\right)^{n-2}
\left(c_n[n-1] +
d_n[n-1]\right) + \left(a \beta_0\right)^{n-2}\beta_0^{-1} \cdot \nonumber\\
\!\!\!\!\!\!\!\!&&\!\!  \Big[d_1 \left(c_{n-1}[n-2] - d_{n-1}[n-2] \right)+ \sum_{m=2}^{n-2}\!\!c_m[m-1]d_{n-m}[n-m-1]
\Big]
                      \bigg\}.
\end{eqnarray}
The first term in (\ref{eq:renormalon}) were presented explicitly in \cite{Broadhurst:1993ru} (Table 2) in notation
 $\frac{4}{3}T_R n_f \to -\beta_0$.
\section{What can we expect for the $\{\beta\}$-expansion in N$^5$LO and beyond}
 \label{sec:4}
Let us consider the general structure of the $\{\beta\}$-expansion starting with the 6 loop result in
Eq.(\ref{eq:d_beta+}) in order $a^5$, $n=5$,
\begin{subequations}
 \label{eq:d_beta+}
\begin{eqnarray}
a^5~~d_5&=& \overbrace{\beta_0^4\, d_5[4]+ \beta_2\beta_0\, d_5[1,0,1]+ \beta_1^2\,d_5[0,2]+\beta_1 \beta_0^2\,d_5[2,1]
+
\beta_3\, d_5[0,0,0,1]}^{p(5-1)}+\nonumber \\
  & &\beta_0^3\, d_5[3]+
     \beta_1\,\beta_0 d_5[1,1]
     + \beta_2\, d_5[0,0,1] + \beta_0^2\, d_5[2]
     + \beta_1\,  d_5[0,1] +\nonumber \\
  & &\beta_0\,d_5[1]+d_5[0]\,,  \label{eq:d_5}    \\
   &\vdots& \nonumber \\
a^n~~ d_{n}
   &=&\underbrace{\overbrace{\beta_0^{n-1}\, d_{n}[n\!-\!1]+ \cdots +\beta_{(n-2)}d_{n}[0,\ldots,1]}^{p(n-1)}+\ldots  +
   d_n[0]}_{N(n)}\,.
\label{eq:d_n}
\end{eqnarray}
\end{subequations}

To obtain the number of elements in this order, we  count them with $ \beta_0^4$  up to $\beta_3$
in the first line of Eq.(\ref{eq:d_5}),
this coincides with the number of partitions $p(5-1)=5$.
The other terms in (\ref{eq:d_5}) repeat the structure of the result in the previous order at $n=4$;
therefore, the complete number of terms at $n=5$ is $(p(5-1)+7)\!=\!\bm{12}$, see the discussion in
\cite{Mikhailov:2016feh}.
Generally speaking, for the term $d_n$ of order of $n$ in Eq.(\ref{eq:d_n}),  one should count new terms from $
\beta_0^{(n-1)}$ up to $\beta_{(n-1)-1}$ that gives their number $p(n-1)$,
whereas the complete number $N(n)$ of all the terms  is evidently the sum $N(n)=\sum_{l=0}^{(n-1)} p(l)$ that leads to
the series
 \begin{eqnarray}
\hspace{-10mm}N(n)\!=\!\!\sum_{l=0}^{(n-1)}p(l)&=& \{1,2,4,7,\bm{12},19,\ldots,97,..\} \sim \frac{\sqrt{6n}}{\pi}\cdot
p(n) +\ldots \,. \\ 
\text{at}~n &=& \{1,2,3, 4,~\bm{5},~~6,\ldots,10,.. \} \nonumber
\end{eqnarray}
Appropriate smooth approximations for $N(n)$ under true asymptotics are presented, \eg, in \cite{OEIS}
and references therein. \\
 The equations for $c_5[.]+d_5[.]$  following from the CBKR are collected in the table below,
the expressions for $c_5[4]$ and  $d_5[4]$ separately are known from \cite{Broadhurst:1993ru}.

\centerline{
\begin{tabular}{|c|c|}
		\hline \hline
		 equations for $c_5[.]+d_5[.]$ from CBKR&Predictions \\
				\hline
	        $c_5[0]+d_5[0]$              & (\ref{eq:c5+d5-b}) \\
		     $c_5[0,0,0,1]+d_5[0,0,0,1]$ & (\ref{eq:c4__1+d4__1})\\
		$c_5[0,0,1]+d_5[0,0,1]$          & (\ref{eq:c3_1+d3_1}) \\
		$c_5[0,1]+d_5[0,1]$              & (\ref{eq:c4_1+d4_1})\\ \hline
			\end{tabular}
}
\vspace*{2mm}

The elements of $c_5 (d_5)$ are formed by a variety of 6-loop diagrams that get contributions from the intrinsic box-
and
pentagon-subgraphs with gluon legs. These subgraphs introduce to $c_5[0] (d_5[0])$ a specific $\Nfr$-dependence  similar
to the one that
appeared for $c_4[0] (d_4[0])$ and that does not relate to the charge renormalization.
Indeed, the new color coefficients (see the definition in Eq.(\ref{dRa1an}))
$ \Ds
~d_A^{a b c d e}d_{F}^{a b c d e}/d_F\,
$ (gluon pentagon inside),
$\Ds \Nfr ~d_{F,r}^{a b c d e}d_{F}^{a b c d e}/d_F
$ (fermion pentagon inside)
enter into $c_5[0]$ together with the contributions from the box-like graphs.
The  $c_5[0]$ can be obtained from $c_5$ following  the scheme discussed for $d_4[0]$ in Sec.\ref{sec:2.1}.
To extract the $c_5[0]$ one can use  the roots  of  the set of equation $\{\beta_{0,1,2,3}(\{
x_{i,0}\})=0\}$,
where variables $\{ x_{0}, x_{1}, x_{2}, x_{3}\}$ are the attributes of d.o.f., see the bold terms in Eq.(\ref{eq:d.o.f.}) in Appendices
\ref{App:A2} and \ref{App:B}.
Then substituting $\{c_{1,2,3,4,5}[0]\}$  in Eq.(\ref{eq:CI-PT0n}) or in Eq.(\ref{eq:c5+d5-b}), one can restore $d_5[0]$.
The element $c_5[0,0,0,1]$ at  $\beta_3$ can then be extracted from the equation
$c_5(x_{0,0},x_{1,0},x_{2,0},x_{3})\!=\!\beta_3(x_{0,0},x_{1,0},x_{2,0},x_{3}) c_5[0,0,0,1]+c_5[0]$,
while its counterpart $d_5[0,0,0,1]$ -- from the prediction in Eq.(\ref{eq:cn_1+dn_1}).
The other elements can be found following the procedure discussed in \cite{Mikhailov:2016feh}
and  briefly sketched in Sec. \ref{sec:2.1}.

A notable attempt to obtain the elements $d_n, c_n$  of the $\{\beta\}$-expansion just for $D$ and $C$
based on the interpretation of the RHS of CBKR  was done in \cite{Goriachuk:2021ayq,Cvetic:2016rot}.
There, firstly, a special ``two-fold'' form of the RHS of CBKR was proposed
that includes one of the  sums over the powers of $\left(\beta(a)/a \right)^j$ and ``works up to N$^{3,4}$LO''.
Secondly, if we assume now such a ``two-fold'' form for each of the factors $D$ and $C$ \textit{separately},
which is the strong \textit{sufficient condition} for CBKR to be satisfied,
one can get all the elements in N$^{3,4}$LO.
Their results for elements of the $\{\beta \}$-expansion for $D, C$ do not coincide with ours in these orders,
while the results for the sums of elements like $c_n[0]+d_n[0]$ in Eq.(\ref{eq:CI-PTsum}) coincides with ours
because their method is already based on CBKR.
The mentioned chain of suggestions allows the authors to obtain 8 elements
out of 12  \cite{Goriachuk:2021ayq} in the N$^5$LO of $\{\beta\}$-expansion
owing to the strength of these  suggestions.
We believe that the method needs further verification.

\section{Conclusion}
 \label{sec:concl}
We have considered here the problem of obtaining the structure of  QCD  corrections  and its elements
by means of the $\{\beta \}$-expansion for the renormalization group invariant quantities: the
nonsinglet parts of the Adler $D$-function and the Bjorken polarized sum rule $C^\text{Bjp}$ --  both in order of
$O(a^4)$.
The explicit results for the elements $d_n[.],~c_n[.]$ of this expansion are obtained in Sec.\ref{sec:2} based on the
extended QCD model with any number of fermion representations (with single coupling constant), QCDe,
which work as new degrees of freedom.
This our approach to constructing the $\{\beta \}$-expansion is universally applied
to any renormalization group invariant quantity \cite{Mikhailov:2016feh}.
Note here that the generalization of the $\{\beta \}$-expansion to the renormalization group
\textit{covariant} quantities was made and discussed in \cite{Kataev:2016aib}.
The explicit knowledge of the elements of the $\{\beta \}$-expansion: \\
(i) gives a possibility to perform various
kinds of optimization of the perturbation series for a variety of important physical quantities,
\eg, related with the Adler $D$-function \cite{Kataev:2014jba};\\
(ii) taken together with the Crewther-Broadhurst-Kataev relation \cite{Crewther:1972kn,Broadhurst:1993ru},
they allow one to establish nontrivial relations between
the afore-mentioned quantities  for high orders (higher than $a^4$) and verify the mutual reliability
of the results in less orders. \\
We show and discuss these relations in Sec.\ref{sec:3} and specially for N$^5$LO in Sec.\ref{sec:4}.
Our results satisfy all of  the above suggested tests.


\acknowledgments

 We would like to thank  K. G. Chetyrkin for essential help,
 useful comments and good advice.
 MS  is thankful to A.~G. Grozin for the important clarifications.

 \appendix
 \section{The $\beta$-function in $O(\alpha_s^4)$ at the extended fermion sector }
\label{App:A}   \setcounter{equation}{0}
\subsection{QCD extended with several fermion representations -- QCDe}
\label{App:A1}
The Lagrangian of a QCD-like model \cite{Zoller:2016sgq,K:2022ebj}, extended to include several
fermion representations of the gauge group is given by
\ba
\ssL_{\sss{QCD}}&=&-\f{1}{4}G^a_{\mu \nu} G^{a\,\mu \nu}-\f{1}{2 \lambda}\lb\p_\mu A^{a\,\mu}\rb^2
+\p_\mu \bar{c}^a \p^{\mu}c^a+\gs f^{abc}\,\p_\mu \bar{c}^a A^{b\,\mu} c^c \nonumber \\
&+&\sum\limits_{r=1}^{N_{\ssst{rep}}}\sum\limits_{q=1}^{\Nfr}
\left\{\f{i}{2}\bar{\psi}_{q,r}\overleftrightarrow{\hat{\p}}\psi_{q,r}-m_{q,r}\bar{\psi}_{q,r}\psi_{q,r}
+ \gs \bar{\psi}_{q,r}\hat{A}^a T^{a,r} \psi_{q,r}\right\}{},
\label{LQCD}
\ea
with the gluon field strength tensor
\be
G^a_{\mu \nu}=\p_\mu A^a_\nu - \p_\nu A^a_\mu + \gs f^{abc}A^b_\mu A^c_\nu{}.
\ee
The index $r$ specifies the fermion representation; and the index $q$, the fermion flavour,
$\psi_{q,r}$ is the corresponding fermion field and $m_{q,r}$ is the corresponding fermion mass.
The number of fermion flavours in the
representation $r$ is $\Nfr$ for any of the $N_{\ssst{rep}}$ fermion representations.

The generators $T^{a,r}$ of each fermion representation $r$ fulfill the defining anticomuting relation of the Lie
Algebra
corresponding to the gauge group:
\be \left[ T^{a,r},T^{b,r} \right]=if^{abc}T^{c,r}\ee
with the structure constants $f^{abc}$.
We have one quadratic Casimir operator $\cfr$ for each fermion representation $r$, defined through
\be T^{a,r}_{ik} T^{a,r}_{kj} = \delta_{ij} \cfr, \ee
and $\ca$ for the adjoint representation. The dimensions of the fermion representations are given by $\dFr$ and the
dimension of the adjoint representation
by $\Ng$. The traces of  different representations are defined as
\be \trr \delta^{ab}=\textbf{Tr}\lb T^{a,r} T^{b,r}\rb=T^{a,r}_{ij} T^{b,r}_{ji}. \ee
At the four-loop level we also encounter higher order invariants in the gauge group factors which are expressed in terms
of
symmetric tensors
\be d_{\sss{R}}^{a_1 a_2 \ldots a_n}=\f{1}{n!}
\sum\limits_{\text{perm } \pi}\text{Tr}\left\{ T^{a_{\pi(1)},R}T^{a_{\pi(2)},R}\ldots T^{a_{\pi(n)},R}\right\}{},
\label{dRa1an}
\ee
where $R$ can be any fermion representation $r$, denoted as $R=\{F,r\}$, or the adjoint representation, $R=A$, where
$T^{a,A}_{bc}=-i\,f^{abc}$.
The standard QCD corresponds to $N_{rep}=1$, while at $N_{rep} > 1$ the first fermion
representation will be considered as a special one -- the standard QCD,
in what follows with
\ba
&&C_{F,1}\equiv C_{F},~d_{F,1}\equiv d_{F},~n_{F,1}\equiv n_{F},~~T^{a,1}\equiv T^{a},~d^{abcd}_{F,1}\equiv
d^{abcd}_{F}\,.
\ea
\subsection{The $\beta$-function in $O(\alpha_s^4)$}
\label{App:A2}
Here we provide the results for the four-loop $\beta$-function of the QCDe coupling $a$ with an arbitrary number of
fermion representations.
The number of active fermion flavours of the representation $i$ are denoted by $\Nfi$,
the Casimir operators -- $\ca$ and $\cfi$
and the traces by $\tri,~d_{\sss{R}}^{a_1 a_2 \ldots a_n}$.
These components form special contributions for d.o.f., those of them that are revealed in
$O(a^4)$ calculations are highlighted in bold below
\ba \label{eq:d.o.f.}
\textbf{nT}&=&\sum\limits_{i} \Nfi \tri,\quad
\textbf{nTC}\bm{k}= \sum\limits_{i} \Nfi \tri  \cfi^k,\quad \textbf{n}\bm{d^{abcd}}= \sum\limits_{i}\Nfi
d^{abcd}_{F,i}\,,
\ea \vspace*{-4mm}

where $\bm{k}$ in \textbf{nTC}$\bm{k}$ is a number.
They enter in
\begin{subequations}
 \label{eq:beta0-3}
\begin{eqnarray}
\beta_{0}&=&
          \f{11}{3} \ca
          -\f{4}{3} \textbf{nT};  \label{1lbetaas}        \\
\beta_{1}&=&  \f{34}{3} \ca^2
          - 4 \left[\textbf{nTC}\bm{1}
          + \f{5}{3} \ca (\textbf{nT})\right];\label{2lbetaas} \\
\beta_{2}&=&  \f{2857}{54} \ca^3
          + 2(\textbf{nTC}\bm{2})
          - \f{205}{9} \ca (\textbf{nT})(\textbf{nTC}\bm{1}) - \f{1415}{27} \ca^2(\textbf{nT})+ \nonumber\\ &&
 \textbf{nT}\left[
         \f{44}{9}(\textbf{nTC}\bm{1})
          + \f{158}{27} \ca (\textbf{nT})\right];\label{3lbetaas}\\
\beta_{3}&=&
          \lb  \f{150653}{486} - \f{44}{9} \zeta_{3} \rb \ca^4
           - \lb \f{80}{9} - \f{704}{3} \zeta_{3} \rb + d_\text{AA}\nonumber\\ 
          &&\left[ 46 (\textbf{nTC3})
          - \lb \f{4204}{27}   - \f{352}{9} \zeta_{3} \rb \ca (\textbf{nTC} \bm{2})
          + \lb \f{7073}{243}  - \f{656}{9} \zeta_{3} \rb \ca^2 (\textbf{nTC}\bm{1})\right.\nonumber\\
          & & \left.
          - \lb \f{39143}{81}  - \f{136}{3} \zeta_{3} \rb \ca^3 (\textbf{nT}) \right]
          + \lb \f{512}{9} - \f{1664}{3} \zeta_{3} \rb \sum\limits_{i} n_{f,i} \,d_{\text{FA},i} + \nonumber\\
%
          && 
          \left[
           \lb \f{184}{3} - 64 \zeta_{3} \rb (\textbf{nTC}\bm{1})^2
          - \lb \f{304}{27}  + \f{128}{9} \zeta_{3} \rb (\textbf{nT}) (\textbf{nTC}\bm{2})  \right.\nonumber\\ & &
          \left.
          + \lb \f{17152}{243}  + \f{448}{9} \zeta_{3} \rb \ca (\textbf{nT}) (\textbf{nTC}\bm{1})
          + \lb \f{7930}{81} + \f{224}{9} \zeta_{3} \rb \ca^2 (\textbf{nT})^2 \right]-    \nonumber\\
          & & \lb \f{704}{9}
          - \f{512}{3}  \zeta_{3} \rb \sum\limits_{i,j}\Nfi \Nfj \,d_{\text{FF},ij}   
\!+\! (\textbf{nT})^2\left[ \f{1232}{243}(\textbf{nTC}\bm{1})
                      + \f{424}{243} \ca (\textbf{nT})\right]\,. \label{4lbetaas}
\end{eqnarray}
 \end{subequations}
where $\Ds T_\text{R} = \frac{1}{2},\, C_\text{F} = \frac{N^2_c-1}{2N_c},\,  C_A =N_c,\, N_\text{A}=2 C_\text{F}
C_\text{A} =N_c^2-1$.
Apart from the mentioned  Casimir operators 
the following invariants appear in our results:
\ba
d_\text{AA} &=& \dAAfNgex,\quad d_{\text{FA},i} = \dFAfiNgex,\quad d_{\text{FF},ij} =\dFFfijNgex, \\
\tilde{d}_\text{FA}&=&\frac{d^{abcd}_{F}d^{abcd}_{A} }{d_F},\quad \tilde{d}_{\text{FF},r} =
\frac{d^{abcd}_{F}d^{abcd}_{F,r} }{d_F}\,,
\ea
where $r$ is fixed and $i,j$ will be summed over all fermion representations.

\section{The Adler D-function and Bjorken SR results in $O(\alpha_s^4)$ at the extended fermion sector}
\label{App:B}
Here we present the results for the coefficients $d_i$ of the Adler $D$-function
obtained in the framework of QCDe in \cite{K:2022ebj}, see Appendix \ref{App:A1}.
\ba
d_1&=& 3\cf\,; \\
d_2&=&-\frac{3}{2}\cf^2+ \cf \ca \left(\frac{123}{2}-44 \zeta_3\right)-2\cf (\textbf{nT})(11-8 \zeta_3)\,;
\\
d_3&=&-\frac{69}{2} \cf^3+ \cf^2\bigg[\ca \left(-127-572  \zeta_3+880 \zeta_5\right)+ (\textbf{nT})(72+ 208
    \zeta_3-320  \zeta_5)\bigg] + \nonumber\\
    &&\cf \ca^2 \left(\frac{90445}{54}-\frac{10948}{9}\zeta_3-\frac{440}{3}\zeta_5\right)+\cf\ca (\textbf{nT})
    \left(-\frac{31040}{27}+\frac{7168}{9} \zeta_3+\frac{160}{3} \zeta_5\right)+\nonumber\\
    &&\cf(\textbf{nT})^2
   \left(\frac{4832}{27}-\frac{1216}{9}\zeta_3\right)+\cf(\textbf{nTC1}) (-101+96 \zeta_3)\,; \\
d_4&=&\cf^4 \left(\frac{4157}{8}+96 \zeta_3\right)+ \nonumber \\
   && \cf^3 \Big[\ca (-2024-278 \zeta_3+18040
   \zeta_5-18480 \zeta_7)-\textbf{nT} (-298 +56 \zeta_3+6560 \zeta_5-6720\zeta_7)\Big]+ \nonumber \\
   &&\cf^2 \bigg[\ca^2\! \left(\!-\frac{592141}{72}\!-\frac{87850}{3}\zeta_3+\frac{104080}{3}\zeta_5+9240
   \zeta_7\right)\!+ \nonumber \\
   &&\phantom{\cf^2 \bigg[}\ca (\textbf{nT})\! \left(\frac{67925}{9}+\frac{61912}{3}\zeta_3-\frac{83680}{3}\zeta_5-3360
   \zeta_7
\right)+\nonumber \\
     &&\phantom{\cf^2 \bigg[}(\textbf{nT})^2
   \left(-\frac{13466}{9}-\frac{10240}{3} \zeta_3+\frac{16000}{3} \zeta_5\right)+
   \textbf{nTC1}
   (251+ 576 \zeta_3-960 \zeta_5)\bigg]+\nonumber
\ea
\ba
   && \cf \Bigg[\ca^3 \left(\frac{52207039}{972}-\frac{912446}{27} \zeta_3-\frac{155990}{9} \zeta_5+4840
   \zeta_3^2-1540
   \zeta_7\right) \nonumber \\
   &&\phantom{\cf \bigg[} \ca^2 (\textbf{nT})
   \left(-\frac{4379861}{81}+\frac{275488}{9}\zeta_3+\frac{150440}{9}\zeta_5-1408
   \zeta_3^2+560 \zeta_7\right)+  \nonumber \\
   &&\phantom{\cf \bigg[}  \ca (\textbf{nT})^2
   \left(\frac{1363372}{81}-\frac{83624}{9} \zeta_3-\frac{43520}{9}\zeta_5-128 \zeta_3^2\right)-  \nonumber \\
   & &\phantom{\cf \bigg[ } \ca (\textbf{nTC1}) \left(2112 \zeta_3^2-7792 \zeta_3-400
   \zeta_5+\frac{375193}{54}\right)\!+\nonumber \\
   &&\phantom{\cf \bigg[}\!(\textbf{nT})^3 \left(-\frac{392384}{243}+
   \frac{25984}{27}\zeta_3 + \frac{1280}{3} \zeta_5\right)+ \nonumber \\
   & &\phantom{\cf \bigg[ }(\textbf{nT}) (\textbf{nTC1})
   \left(\frac{63250}{27}-2784 \zeta_3+768 \zeta_3^2\right)+\textbf{nTC2} \left(\frac{355}{3}+
   272 \zeta_3-480 \zeta_5\right)\Bigg]-
   \nonumber \\
   && 16\left[\sum\limits_{r} n_{f,r} \,\tilde{d}_{\text{FF},r}\cdot\left(13+ 16\zeta_3-
   40\zeta_5\right)+\tilde{d}_\text{FA}\cdot\left(-3+ 4\zeta_3+20\zeta_5\right)\right]\,.
     \ea
The results for $c_k$ of the Bjorken SR in QCDe,
\ba
c_1&=& -3\cf\,; \\
c_2&=&\frac{21}{2} \cf^2 - 23 \ca \cf + 8 \cf (\textbf{nT})\,;
\ea
 \ba
c_3&=&-\frac{3}{2} \cf^3+
\cf^2 \left[\ca
   \left(\frac{1241}{9}-\frac{176}{3} \zeta_3\right)-\textbf{nT}
   \left(\frac{664}{9}-\frac{64}{3}\zeta_3\right)\right]+\cf\ca^2 \left(-\frac{10874}{27}+\frac{440}{3}\zeta_5\right)+
   \nonumber \\
 && \cf\ca(\textbf{nT}) \left(\frac{7070}{27}+48 \zeta_3-\frac{160}{3} \zeta_5\right)
 -\cf (\textbf{nT})^2\frac{920}{27}+ \cf (\textbf{nTC1}) (59-48 \zeta_3); \\
c_4&=&-\cf^4 \left(\frac{4823}8 + 96 \zeta_3\right)  +\nonumber \\
    &&\phantom{-}\cf^3 \left[ -
    \ca \left(\frac{3707}{18} + \frac{7768}3 \zeta_3 - \frac{16720}3 \zeta_5\right)+
    \textbf{nT} \left(\frac{5912}{9} + \frac{3296}3 \zeta_3 - \frac{6080}3 \zeta_5\right)\right]+ \nonumber \\
 && \phantom{-}\cf^2 \bigg[
    \ca^2 \left(\frac{1071641}{216} + \frac{25456}{9} \zeta_3 - \frac{22000}{9} \zeta_5 - 6160 \zeta_7\right) -\nonumber
    \\
 &&\phantom{ -\cf^2 \bigg[ }   \ca (\textbf{nT}) \left(\frac{106081}{27} + \frac{9104}{9} \zeta_3 - \frac{8000}{9}
 \zeta_5 - 2240 \zeta_7\right)+ \nonumber \\
 &&\phantom{ -\cf^2 \bigg[ } (\textbf{nT})^2 \left(\frac{16114}{27} - \frac{512}{3} \zeta_3\right) - \textbf{nTC1}
    \left(\frac{1399}{3} - 400 \zeta_3\right)\bigg] + \nonumber
 \ea
\ba
 && \cf \Bigg[ \ca^3 \left(-\frac{8004277}{972} + \frac{4276}{9} \zeta_3 + \frac{25090}{9} \zeta_5
    - \frac{968}{3} \zeta_3^2 +
       1540 \zeta_7\right) + \nonumber \\
&&\phantom{\cf \Bigg[}   \ca^2 (\textbf{nT}) \left(\frac{1238827}{162} + 236 \zeta_3 - \frac{14840}{9} \zeta_5 +
\frac{704}{3} \zeta_3^2-
       560 \zeta_7\right) - \nonumber \\
 &&\phantom{\cf \Bigg\{}
       \ca(\textbf{nT})^2 \left(\frac{165283}{81} + \frac{688}{9} \zeta_3 - \frac{320}{3} \zeta_5+ \frac{128}{3}
       \zeta_3^2\right)+\nonumber \\
 &&\phantom{\cf \Bigg\{} \ca (\textbf{nTC1}) \left(\frac{124759}{54} - 1280 \zeta_3 - 400 \zeta_5\right)+\nonumber \\
 && \frac{38720}{243} (\textbf{nT})^3 - (\textbf{nT})(\textbf{nTC1}) \left(\frac{19294}{27} - 480 \zeta_3\right)
    -\textbf{nTC2} \left(\frac{292}3 + 296 \zeta_3 - 480 \zeta_5\right)
       \Bigg] + \nonumber \\
  & & 16\left[
       \sum\limits_{r} n_{f,r} \,\tilde{d}_{\text{FF},r}\cdot\left(13+ 16\zeta_3-
   40\zeta_5\right)+\tilde{d}_\text{FA}\cdot\left(-3+ 4\zeta_3+20\zeta_5\right)
       \right]\,.
\ea


\begin{thebibliography}{20}
\bibitem{Crewther:1972kn}
 R.~J.~Crewther,
 ``Nonperturbative evaluation of the anomalies in low-energy theorems,''
 Phys.\ Rev.\ Lett.\  {\bf 28} (1972) 1421.

\bibitem{Broadhurst:1993ru}
  D.~J.~Broadhurst and A.~L.~Kataev,
  ``Connections between deep inelastic and annihilation processes at next to
  next-to-leading order and beyond,''
  Phys.\ Lett.\  B {\bf 315} (1993) 179.

\bibitem{Mikhailov:2004iq}
  S.~V.~Mikhailov,
  ``Generalization of BLM procedure and its scales in any order of pQCD: A Practical approach'',
  JHEP \textbf{2007}, 009 (2007),
  doi: 10.1088/1126-6708/2007/06/009,
  [hep-ph/0411397].

\bibitem{Mikhailov:2016feh}
  S. V.  Mikhailov,
    ``On a realization of $\{\beta\}$-expansion in QCD''
      JHEP \textbf{2017}, 169 (2017),
  doi: 10.1007/JHEP04(2017)169,
            [arXiv:1610.01305].

\bibitem{Kataev:2010du}
  A.~L.~Kataev and S.~V.~Mikhailov,
 ``Conformal symmetry breaking effects in gauge theories: new perturbative expressions'',
  Theor. Math. Phys. \textbf{ 170} (2012) 139,
doi: 10.1007/s11232-012-0016-7,
[arXiv:1011.5248].

\bibitem{Kataev:2014jba}
  A.~L.~Kataev and S.~V.~Mikhailov,
  ``Generalization of the Brodsky-Lepage-Mackenzie optimization within the $\{\beta\}$-expansion and the principle of
  maximal conformality'',
  Phys.\ Rev. \textbf{D 91}, 014007 (2015),
doi: 10.1103/PhysRevD.91.014007,
  [arXiv:1408.0122].

\bibitem{Kataev:2016aib}
 A.~L.~Kataev and S.~V.~Mikhailov,
``The $\{\beta\}$-expansion formalism in perturbative QCD and its extension'',
JHEP\textbf{2016}, 079 (2016),
doi: 10.1007/JHEP11(2016)079,
[arXiv:1607.08698].

\bibitem{Baikov:2010je}
     P. A. Baikov, K. G. Chetyrkin, J. H. Kuhn,
     ``Adler Function, Bjorken Sum Rule, and the Crewther
                        Relation to Order $\alpha_s^4$ in a General Gauge Theory'',
     Phys. Rev. Lett. \textbf{104}, 132004 (2010),
      	[arXiv:1001.3606]\,.




\bibitem{Baikov:2012zn}
 P. A. Baikov and K. G. Chetyrkin and J. H. Kuhn and J. Rittinger,
    ``Adler Function, Sum Rules and Crewther Relation of Order $\mathcal{O}(\alpha^4_s)$: the Singlet Case'',
    Phys. Lett. \textbf{B}714 (2012) 62, [arXiv:1206.1288]\,.


\bibitem{Brodsky:1982gc}
  S.~J.~Brodsky, G.~P.~Lepage and P.~B.~Mackenzie,
    ``On the Elimination of Scale Ambiguities in Perturbative Quantum Chromodynamics'',
  Phys.\ Rev. \textbf{ D 28}, 228 (1983).

\bibitem{Zoller:2016sgq}
   M. F.  Zoller,
     ``Four-loop QCD $\beta$-function with different fermion
                        representations of the gauge group'',
JHEP \textbf{2016}, 118 (2016),
doi: 10.1007/JHEP10(2016)118,
  [arXiv:1608.08982]\,.

\bibitem{Chetyrkin:2017mwp}
K. G. Chetyrkin and M. F. Zoller,
``Four-loop renormalization of QCD with a reducible fermion representation of the gauge group: anomalous dimensions and
renormalization constants'',
JHEP\textbf{2017}, 074 (2017),
 doi: 10.1007/JHEP06(2017)074,
 [arXiv:1704.04209]\,.

\bibitem{K:2022ebj}
K. G. Chetyrkin,
``Adler function, Bjorken Sum Rule and
Crewther-Broadhurst-Kataev relation with generic
fermion representations at order $O(\alpha_s^4)$'',
 arXiv: 2206.12948 [hep-ph],
 doi:10.48550/ARXIV.2206.12948\,.

\bibitem{Crewther:1997ux}
R.~J.~Crewther,
``Relating inclusive $e^{+} e^{-}$ annihilation to electroproduction
sum rules in Quantum Chromodynamics''
Phys. Lett. \textbf{B 397} (1997) 137,
doi: 10.1016/S0370-2693(97)00157-3,
  [hep-ph/9701321].

\bibitem{Braun:2003rp}
 V.~M.~Braun, G.~P.~Korchemsky, Dieter~M\"uller,
``The Uses of conformal symmetry in QCD''
Prog. Part. Nucl. Phys. \textbf{51}, 311, 2003,
doi: 10.1016/S0146-6410(03)90004-4,
[hep-ph/0306057].

\bibitem{Broadhurst:1992si}
D.~J.~Broadhurst,
``Large N expansion of QED: Asymptotic photon propagator and contributions to the muon anomaly, for any number of
loops'',
Z. Phys. C \textbf{58} (1993), 339-346
doi:10.1007/BF01560355

\bibitem{Chetyrkin:1996ez}
  K.~G.~Chetyrkin,
 ``Corrections of order alpha-s**3 to R(had) in pQCD with light gluinos'',
  Phys.\ Lett.\ \textbf{B 391} (1997) 402
   [hep-ph/9608480].


\bibitem{Clavelli:1996pz}
  L.~Clavelli, P.~W.~Coulter and L.~R.~Surguladze,
``Gluino contribution to the 3-loop beta function in the minimal
  supersymmetric standard model'',
  Phys.\ Rev.\ \textbf{ D 55} (1997) 4268.

\bibitem{Ma:2015dxa}
H.~H.~Ma, X.~G.~Wu, Y.~Ma, S.~J.~Brodsky and M.~Mojaza,
``Setting the renormalization scale in perturbative QCD: Comparisons of the principle of maximum conformality with the
sequential extended Brodsky-Lepage-Mackenzie approach,''
Phys. Rev. D \textbf{91} (2015), 094028
doi:10.1103/PhysRevD.91.094028


\bibitem{Cvetic:2016rot}
       G. Cvetic and  A. L. Kataev,
 ``Adler function and Bjorken polarized sum rule:
Perturbation expansions in powers of the SU(N$_c$)
                        conformal anomaly and studies of the conformal symmetry
                        limit'',
    Phys. Rev.\textbf{D 94},
    1, 014006,
  [arXiv:1604.00509].

\bibitem{grozin:2007}
A.~G. Grozin, ``\textit{{QED and QCD}}'',  in ``\textit{Lectures on QED and QCD}'', ch.~1.3.9,
\newblock {World Scientific Publishing Co. Pte. Ltd.}, 1~ed., 2007,
doi:10.1142/6200.


\bibitem{OEIS}
``The On-Line Encyclopedia of Integer Sequences'', https://oeis.org/A000070\,.

\bibitem{Goriachuk:2021ayq}
I. O. Goriachuk and A. L. Kataev and V. S. Molokoedov,
``The ${\rm{\overline{MS}}}$-scheme $\alpha_s^5$ QCD contributions to the Adler function and Bjorken polarized sum rule
in the Crewther-type two-fold $\{\beta\}$-expanded representation'',
JHEP\textbf{2022}, 028 (2022),
doi: 10.1007/JHEP05(2022)028,
[arXiv:2111.12060].
\end{thebibliography}
\end{document}